\documentclass[aps,twocolumn,showpacs,prl]{revtex4}
\usepackage{graphicx}
\usepackage{dcolumn}
\usepackage{amsmath,amssymb,graphicx,bm,epsfig}
\usepackage{epstopdf}

\renewcommand{\a}{\alpha}
\renewcommand{\b}{\beta}

\begin{document}
\draft \preprint{J. Kim $et~al.$}

\title {Termination-dependent Surface In-gap States in a Mixed-valent Topological Insulator:  SmB$_6$}
\author{Junwon Kim$^1$, Kyoo Kim$^1$, Chang-Jong Kang$^1$, Sooran Kim$^1$, Hong Chul Choi$^1$, 
J.-S. Kang$^2$, J. D. Denlinger$^3$, and B. I. Min$^1$}
\email[e-mail: ]{bimin@postech.ac.kr}
\affiliation{
$^1$Department of Physics, PCTP,
        Pohang University of Science and Technology,
        Pohang 790-784, Korea\\
$^2$Department of Physics, The Catholic University of Korea,
        Bucheon 420-743, Korea\\
$^3$Advanced Light Source, Lawrence Berkeley Laboratory,
        Berkeley, CA 94720, U.S.A.
}
\date{\today}

\begin{abstract} 
{
We have investigated the surface states of a potential mixed-valent topological insulator SmB$_6$
based on the first principles density functional theory slab band structure analysis.
We have found that metallic surface states are formed in the bulk band gap region,
providing evidence for the topological insulating nature of SmB$_6$.
The obtained surface in-gap states are quite different from those in existing reports in that
they are formed differently depending on the Sm or B$_6$ surface termination, 
and are composed of mainly Sm $4f$ state indicating the essentiality of including $f$ electrons 
in describing the surface states.
We have obtained the spin chiral structures of the Fermi surfaces,
which are also in accordance with the topological insulating nature of SmB$_6$.
}
\end{abstract}


\maketitle


The low-temperature residual conductivity in a typical Kondo insulator 
SmB$_6$ has been a long-standing unresolved problem.
In this respect, recent reports on the topologically protected metallic surface states
in SmB$_6$ have attracted great attentions.
\cite{transport_Fisk, Shi-ARPES, Hasan-ARPES, MinChulHee-ARPES, 
Denlinger-ARPES, Wang-STS, Chen-AMR, Coleman1, Coleman2, Takimoto, Gutzwiller}
Angle-resolved photoemission spectroscopy (ARPES) can provide direct evidence for
topological insulators via observation of the Fermi surfaces 
arising from the surface in-gap states.
Indeed, Fermi surfaces were observed in several ARPES studies 
at $\bar{\Gamma}$ and $\bar{X}$ points of the surface Brillouin zone (BZ), 
\cite{Shi-ARPES, Hasan-ARPES, Golden-ARPES,Kimura-ARPES}
supporting the topological Kondo insulating nature of SmB$_6$.
Unfortunately, however, no explicit Dirac cone feature has been observed.   
Hence, some groups interpreted the metallic states observed in ARPES 
not as topological surface states but as bulk-shifted or normal (trivial) surface states.
\cite{Pol_driven_SS, Golden-ARPES}
Especially, the origin of the metallic bands at $\bar{\Gamma}$ and $\bar{X}$ is controversial
as to whether they come from bulk or surface states.
\cite{Shi-ARPES, Hasan-ARPES, Denlinger-ARPES, Golden-ARPES, Kimura-ARPES}
Above all, the spin polarizations of the metallic surface bands has not been unveiled yet.
\cite{MinChulHee-ARPES,Imada-ARPES}  

Among many difficulties, the surface instability in SmB$_6$ 
prevents revealing the topological nature.
Since the chemical bonding between Sm and B$_6$ ions is not weak,
the natural cleavage plane in SmB$_6$ is not well defined.
The pristine Sm- and B$_6$-terminated surfaces, 
which are electrically polar and highly reactive,
are likely to undergo disordered reconstructions or contaminations.
Due to this reason, the topological nature arising from the intrinsic surface states 
is not easily confirmed experimentally.
Another difficulty with SmB$_6$ is the strong correlation effect of $f$ electrons.
The highly renormalized $f$ bands require extremely high-resolution ARPES experiment.
Nevertheless, the topological invariance of SmB$_6$, 
which has been confirmed by a couple of theoretical studies,\cite{Takimoto,Gutzwiller}
arouse great curiosity about the characteristics of the metallic surface states
in SmB$_6$.\cite{Robustness}

There have been a few surface band calculations 
which tried to identify the topological properties of SmB$_6$.
Takimoto\cite{Takimoto} and Lu {\it et al.}\cite{Gutzwiller} performed the (001) surface 
model calculations based on the density functional theory (DFT) 
and the DFT+Gutzwiller bulk band structures, respectively.
Both of them obtained the metallic surface states in the bulk gap region, which produce 
three Dirac cones and the corresponding Fermi surfaces at $\bar{\Gamma}$ and $\bar{X}$.
This feature suggests the nontrivial topological nature of SmB$_6$.
But they considered neither the surface termination dependence nor the surface relaxation effect.
Zhu {\it et al.}\cite{Pol_driven_SS} performed the DFT slab calculations, 
and also obtained the surface states in the vicinity of the Fermi level (E$_F$).
But their surface states 
come from the polarity-driven boron dangling bond, and so they claimed that 
the metallic surface states in SmB$_6$ are not the topologically protected surface states
but the normal surface states.
In their surface slab calculations, however, they considered the Sm $f$ electrons as core.
Since the $f$-$d$ hybridization is essential to develop the insulating state in bulk SmB$_6$, 
it is likely that their surface band structures are not relevant to $f$-electron system of SmB$_6$.

In this letter, we have investigated systematically the (001) surface states 
of a potential mixed-valent topological insulator SmB$_6$,
using the first principles DFT calculations 
on the slabs with different surface terminations.
We have first compared the bulk band structures 
obtained by the DFT with those obtained by the dynamical mean-field theory (DMFT),
whereby we have shown that the DFT is useful to investigate the 
low energy band structure of strongly correlated mixed-valent insulator SmB$_6$.
Then, we have performed the DFT surface band structure calculations, and found that 
the surface bands and the corresponding Fermi surfaces 
are quite different from those in literature.\cite{Takimoto,Gutzwiller,Pol_driven_SS} 
The gapless surface states are formed differently for the Sm- and B$_6$-terminated surfaces.
Moreover, additional surface states appear around $\bar{M}$ for the Sm-terminated case.
We thus argue the importances of considering the Sm $f$ electrons and 
the termination dependence in describing the surface states of SmB$_6$.
Further, we have carefully examined the spin chiralities of the Fermi surfaces 
to corroborate the topological insulating nature of SmB$_6$.

\begin{figure}[t]
\begin{center}
\includegraphics[width=0.35\textwidth]{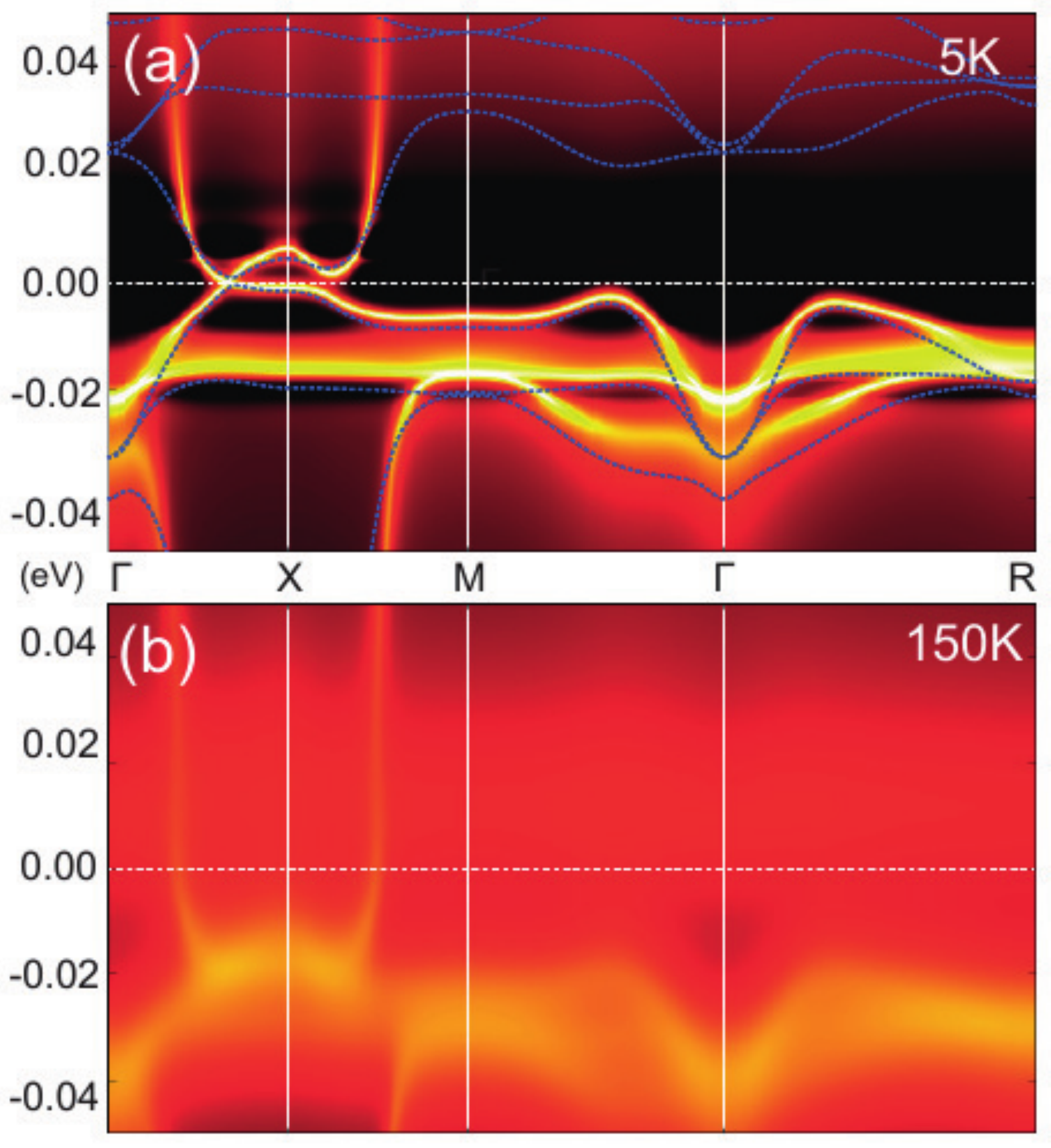} 
\caption{
(Color online)
 Momentum-resolved spectral function of SmB$_6$ obtained by the DFT+DMFT method
 at (a) T=5 K and (b) T=150 K.
Transition to the metallic state is clearly visible at higher temperature 150K.\cite{Denlinger-ARPES}
{\bf
}
The GGA band structure (blue solid lines), 
which is rescaled by 1/10, is overlaid on the DMFT band structure.
}
\label{dmft}
\end{center}
\end{figure}


We have investigated the electronic structures of bulk SmB$_6$   
using both the DFT and the DMFT incorporating the spin-orbit coupling (SOC). 
To describe the surface states, two slab geometries were used ;
Sm- and B$_6$-terminated slabs.
For the DFT, we employed both WIEN2k\cite{WIEN2k} and VASP code.\cite{VASP},
while, for the DFT+DMFT, we used the scheme implemented based on WIEN2k.\cite{DMFT}
To scrutinize the spin chirality of each Fermi surface originating from 
the surface states, the spin-noncollinear calculations were performed.
Further details are provided in the supplement.\cite{Supp}
%


DMFT band structures in Fig.~\ref{dmft} demonstrate the temperature (T) dependent
$f$-$d$ hybridization behavior.
At high T=150K in Fig.~\ref{dmft}(b), Sm $f$ electrons do not form coherent bands,
and so the metallic Sm $d$ band crossing E$_F$ is clearly visible around  X,
reflecting little hybridization with incoherent $f$-bands.
Upon cooling, the $f$ electrons form the coherent bands
that become strongly hybridized with the $d$ band so as to display a gap feature,
even though the gap size is almost zero in the DMFT result of Fig.~\ref{dmft}(a) (T=5 K).


In Fig.~\ref{dmft}(a), the DMFT bands are overlaid with DFT bands 
that are rescaled down in energy by 1/10 (thin dotted blue).
It is seen that the DMFT and DFT band structures near E$_F$ are 
essentially the same as each other. 
This feature suggests that the strong correlation effect of $f$ electrons can be captured 
to some extent just by renormalizing DFT bands.
The gap feature persists in the renormalized DFT band structure with a reduced size of $\sim 2$ meV
with respect to the original gap size of $\sim 20$ meV.
Of course, the renormalized DFT cannot simulate the Sm $j=7/2$ bands that are to be shifted up
by the strong correlation of Sm $4f$ electrons. 
So the position of Sm $j=7/2$ bands in the DFT is located  much closer to E$_F$ than in the DMFT.
In fact, the present authors have previously analyzed the influence 
of the low-lying Sm $j=7/2$ bands on the states near E$_F$.\cite{CJ}
This effect is discussed more in the supplement.\cite{Supp}
\begin{figure}[t]
\begin{center}
\includegraphics[width=0.39\textwidth] {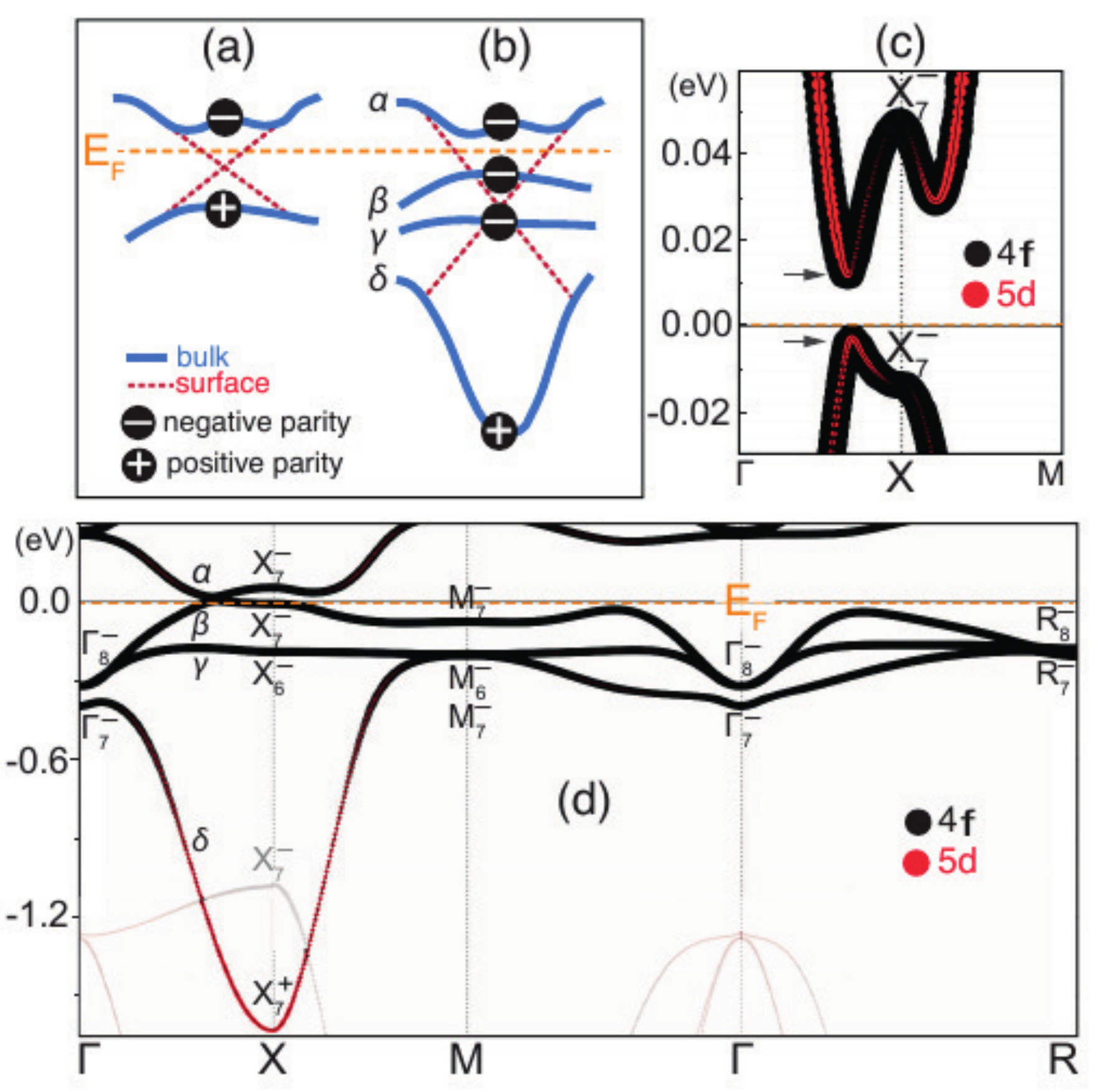} 
\caption{
(Color online)
Schematic pictures showing (a) usual and (b) unusual band inversions. 
In SmB$_6$, parity-inverted bands are $\a$ and $\delta$,
but two more bands $\b$ and $\gamma$ are located between them.
(c) In SmB$_6$, the DFT bands just below and above E$_F$ do not show band inversion at $X$,
because both of them have the $f$-orbital character having the negative parity.
Black and red dots represent the weights of Sm $4f$ and $5d$ components, respectively.
(d) DFT bulk band structure of bulk SmB$_6$.
As shown in (b), two more bands with negative parity are located below E$_F$.
%
}
\label{binv}
\end{center}
\end{figure}

\begin{figure*}[t]
\begin{center}
\includegraphics[width=0.90\textwidth] {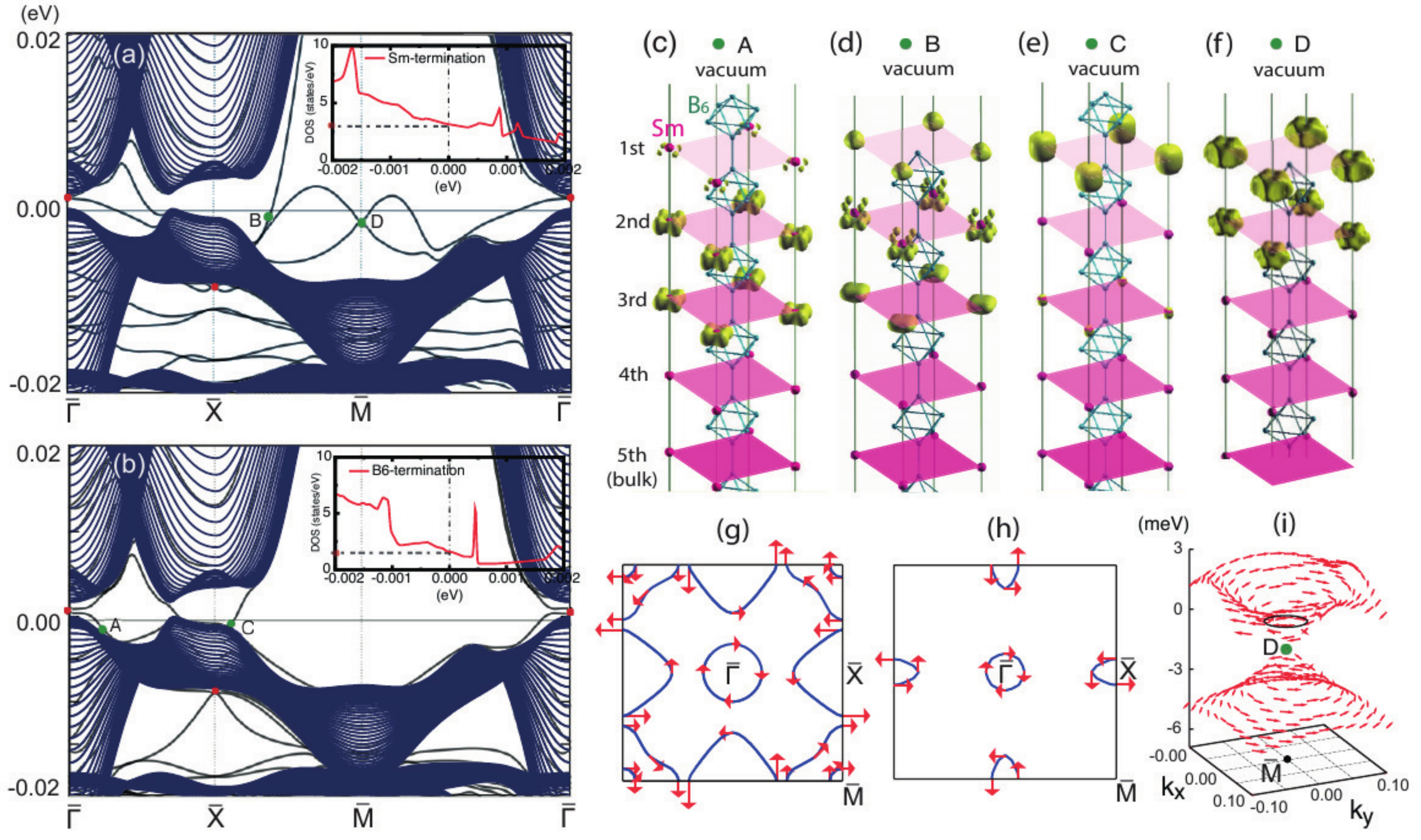} 
\caption{
(Color online) 
DFT band structures of SmB$_6$ slabs with 
(a) the Sm-terminated surface and (b) the B$_6$-terminated surface.
{\bf
}
Note that the energy scales in (a) and (b) are reduced by 1/10.
{\bf
}
Blue shaded region represents the projection of SmB$_6$ bulk bands 
to the (001) surface BZ.
Red dots represent the crossing (Dirac) points of metallic surface bands.
Insets show the DOS (per Sm layer) of each surface termination.
(c)-(f) Charge densities of wave functions at given points A, B, C, and D (green dots)
of the metallic surface bands.
(g),(h) Fermi surfaces of the Sm- and the B-termination, respectively.
Spin helicity in each Fermi surface is plotted with red arrow.
(i) Dirac cone-like shape of the spin chiral texture of the surface state at $\bar{M}$.
}
\label{sband}
\end{center}
\end{figure*}


In typical topological insulators such as Bi$_2$Se$_3$ or Bi$_2$Te$_3$, 
the parity inversion occurs at a high symmetry point  
between two bulk bands having different parities, 
and the topologically protected surface states emerge in the gap region,
as shown in Fig.~\ref{binv}(a).
SmB$_6$, however, shows peculiar parity inversion feature.
As discussed in Fig.~\ref{dmft},
the occurrence of the gap in SmB$_6$ is initiated by the $f$-$d$ hybridization near X,
but the band gap is seen to be realized between two $f$-bands, as shown Fig.~\ref{binv}(b)-(d).
Namely, there are two more $f$ bands, $\b$ and $\gamma$, having the same negative parity (X$_7^{-}$)
in-between parity inverted bands of $\a$ (X$_7^{-}$) and $\delta$ (X$_7^{+}$).
Therefore, some part of the surface in-gap states including the Dirac point  
would be buried in the bulk $\b$ and $\gamma$ bands, as in Fig.~\ref{binv}(b).
This feature in SmB$_6$ is different from usual topological insulators,
in which the Dirac cone features are clearly observed in the gap region.
This is an important difference, making SmB$_6$ more complicated than 
other conventional topological insulators.


In order to probe into the surface states of SmB$_6$, 
we have performed the DFT slab calculations 
for both Sm- and B$_6$- terminated (001) surfaces,
including Sm $f$ electrons as valence electrons.
Figure~\ref{sband}(a) and (b) show the surface states of 
Sm- and B$_6$- terminated SmB$_6$ slabs, respectively.
In both cases, there appear metallic surface states in the bulk gap region,
which are seen to merge into bulk bands projected to the surface BZ.  
The surface in-gap states here are composed of mainly Sm $f$ state
(the weight of $f$-component is more than 90\%).
This result is quite contrary to the previous DFT surface band calculation,\cite{Pol_driven_SS}
in which the surface states near E$_F$ came mostly from B-$2p$ dangling bond state. 
This discrepancy arises from that Sm $f$ electrons were not taken into account 
as valence electrons in Ref.~\cite{Pol_driven_SS}.
Without Sm-$f$ electrons, we also obtained the B-$2p$ surface states that cross E$_F$.\cite{2.0SS} 
{\bf 
}
However, in this case, there are many other metallic bands near E$_F$ besides the B-$2p$ surface bands
due to the lack of hybridization with $f$ electrons, 
which is certainly in disagreement with ARPES data.
Note that the B-$2p$ surface states do not satisfy the criterion of the topological insulator,
whereas the present surface states including Sm $f$ electrons do on both terminations.
Thus the inclusion of $f$ electrons as valence state changes the situation near E$_F$ dramatically.


At a glance, the surface states in Fig.~\ref{sband}(b) for the B$_6$-terminated case
look analogous to
those obtained by the tight-binding (TB) surface model calculation based on 
the LDA + Gutzwiller bulk band result.\cite{Gutzwiller}
However, there are interesting differences.
First, in the TB result, the gapless surface states,
especially at $\bar{X}$, seem to be generated from the band $\b$
manifesting the Dirac points in the gap region,
while, in the present rescaled DFT, the Dirac point is seen to be just buried in the band $\b$.
This is not a minor distinction. 
Since the parities of bulk band $\b$ and $\gamma$ are negative,
the absence of both bands does not change the topological order.
So the gapless surface states are not formed via the bulk band $\b$ and $\gamma$,
but formed just passing through both bands.\cite{Soo_yong}
Interestingly, a similar feature of buried Dirac point in the bulk bands is also seen in PuB$_6$, 
which was reported to be a correlated topological insulator.\cite{PuB6}

Secondly, Sm-termination in Fig.~\ref{sband}(a) 
shows additional surface in-gap states centered at $\bar{M}$,
which have never been recognized theoretically.
The existence of these metallic surface states does not violate the E$_F$ crossing criterion
of Z$_2$ topological characterization.
The different surface states depending on the terminations are quite natural 
even in conventional topological insulators.\cite{termination_dep}
It is, however, rarely reported in real materials that additional metallic surface states 
emerge with different dispersions at another high symmetry point.
Metallic surface states centered at $\bar{M}$ have not been reported experimentally either. 
{\bf
%
}
One reason why $\bar{M}$-centered surface states have not been detected
might be due to a fact that 
the size of well-ordered $1 \times 1$ Sm-terminated surface known until now
is not over a few nanometers.\cite{Harvard-STS,DRESDEN-STS}
The beam spot size of state-of-art ARPES 
may not be enough to capture the physics of such small $1 \times 1$ ordered
Sm-terminated surface.


According to recent STM/STS\cite{DRESDEN-STS} data
on the non-reconstructed ordered Sm- and B$_6$-surface terminated SmB$_6$,
the differential conductance of the former is about 30$\%$
larger than that of the latter at T=4.6 K.
This implies that there are additional conducting channels at the Sm-terminated surface,
suggesting the higher density of states (DOS) near E$_F$.
Indeed, insets of Fig.~\ref{sband}(a) and(b) shows the surface DOSs, 
which shows about 50$\%$ larger DOS
at E$_F$ in the case of Sm-terminated surface,
which matches well with the STS/STM result.
So the metallic surface states at $\bar{M}$ can explain
the larger differential conductance on the Sm-terminated surface.


Figure~\ref{sband}(c)-(f) show the charge densities of surface wave functions at specific
${\bf k}$ points.
It is shown that the charge densities of the surface states originate indeed 
from the surface atoms, mainly from Sm atoms, of both terminated slabs.
It is worthwhile to note in Fig.~\ref{sband}(c) that 
the surface state near $\bar{\Gamma}$ (A) is quite discernible from other surface states 
near $\bar{X}$ and $\bar{M}$ (B, C, D),
in that the dominant charge density at A is slightly away from the topmost layer
and locates rather close to the bulk side.
This is also the case for the surface state at $\bar{\Gamma}$ for the Sm-termination.
This feature suggests that the surface state at $\bar{\Gamma}$ could be robuster 
than other surface states against any changes at the surface.


Figure~\ref{sband}(g) and (h) present the Fermi surfaces (FSs) 
of surface states together with spin chiral structures.
It is seen that both terminations have FSs at $\bar{\Gamma}$ and $\bar{X}$ 
with the same helical spin polarization,  
even though the FSs of the B$_6$-termination are smaller than those of the Sm-termination.
In both cases, FSs at $\bar{\Gamma}$ and $\bar{X}$ are hole and electron FSs, respectively.
Also, due to the surface states of the Sm-termination around $\bar{M}$,
there appear additional FSs of flower shape centered at $\bar{M}$ for the Sm-termination case
in Fig.~\ref{sband}(g). 
These FSs also have helical spin polarizations.
The spin texture of the surface state at $\bar{M}$ is more specifically shown 
in Fig.~\ref{sband}(i), which reveals the opposite helicities 
above and below the Dirac point D.
Therefore, not only previously reported surface states at $\bar{\Gamma}$ and $\bar{X}$
but also that at $\bar{M}$ has Dirac cone-like spin texture.  
This feature corroborates the topological nature of the surface states in SmB$_6$. 
One thing to be pointed out is that the $\bar{M}$-centered FSs could be fragile
with respect to any environmental perturbations on the surface.
It is because, even without the $\bar{M}$-centered FSs,
the criterion for nontrivial topology, {\it i.e.} an odd number of E$_F$ crossings, 
is still valid.\cite{Supp}




In conclusion, we have investigated the surface in-gap states of SmB$_6$ based on the 
first principles DFT slab band structure analysis.
We have found that 
(i) the surface states and corresponding FSs emerge at around $\bar{\Gamma}$ and $\bar{X}$,
(ii) there appear additional surface states centered at $\bar{M}$ for the Sm-termination case,
and (iii) more importantly, the present surface states are quite different from those 
of existing surface band calculations,
such as the TB slab calculation and the DFT calculation without considering Sm 4$f$ electrons.
We have also determined the spin helicities of Fermi surfaces,
which are consistent with the topological Kondo insulating nature of SmB$_6$. 
For the clear confirmation, spin-resolved ARPES on the ordered surface 
is strongly recommended.

Acknowledgments$-$
We would like to thank Jinwoong Kim and  J. H. Shim for helpful discussions.
This work was supported by the NRF (Grant No.2009-0079947, No. 2011-0025237),
the POSTECH BSRI Grant, and the KISTI supercomputing center (Grant No. KSC-2012-C3-055).
J.S.K. acknowledges support by the NRF (Grant No. 2011-0022444).
J.D.D. is supported by the U.S. DOE (Grant No. DE-AC02-05CH11231).




\begin{thebibliography}{}
\bibitem{2.0SS} The additional B-$2p$ surface states at about $-2.0$ eV for the Sm-terminated case
	are not influenced by the inclusion of the $f$ electrons.
\end{thebibliography}
\end{document}


\title {Supplemental Material : \\
Termination-dependent Surface In-gap States in a Mixed-valent Topological Insulator: SmB$_6$}
\author{Junwon Kim$^1$, Kyoo Kim$^1$, C. -J. Kang$^1$, Sooran Kim$^1$, H. -C. Choi$^1$, 
J.-S. Kang$^2$, J. D. Denlinger$^3$, and B. I. Min$^1$}
\affiliation{
$^1$Department of Physics, PCTP,
        Pohang University of Science and Technology,
        Pohang 790-784, Korea\\
$^2$Department of Physics, The Catholic University of Korea,
        Bucheon 420-743, Korea\\
$^3$Advanced Light Source, Lawrence Berkeley Laboratory,
        Berkeley, CA 94720, U.S.A.
}
\date{\today}
\maketitle

\begin{figure*}[b]
\begin{center}
\includegraphics[width=0.80\textwidth]{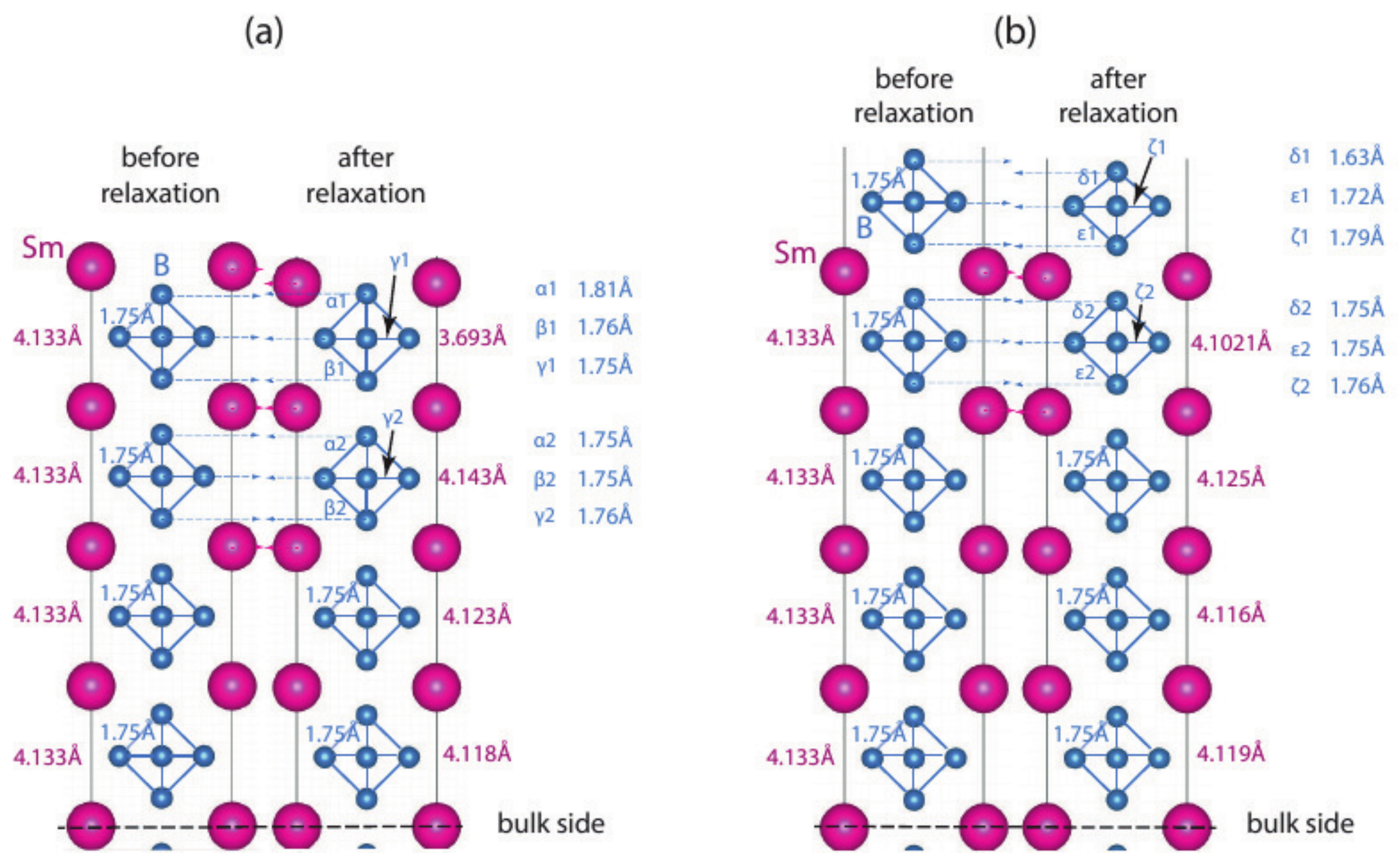}
\caption{
(Color online)
Relaxed structures of SmB$_6$ slabs having (a) Sm- and (b) B$_6$-terminated surfaces.
After the structural relaxation, the thicknesses of both slabs are reduced.
}
\label{rlx}
\end{center}
\end{figure*}


\section{Computational Details and Structural Relaxation}
SmB$_6$ crystallizes in a cubic structure of CaB$_6$-type with space group 
\textit{Pm}3\textit{m} (No.221) and the lattice constant of 4.133 $\AA$.
We have investigated the electronic structures of bulk SmB$_6$   
using both the DFT and the DMFT incorporating the spin-orbit coupling (SOC). 
we used the DFT+DMFT scheme,\cite{DMFT} implemented based on WIEN2k.\cite{WIEN2k}
Coulomb and exchange interaction parameters were set by 7.00 eV and 0.83 eV, respectively.  
For the impurity solver of the DMFT, we used both the non-crossing approximation (NCA) 
and the one-crossing approximation (OCA).\cite{DMFT}
Meanwhile, to describe the surface states, 
two non-stoichiometric supercell geometries were employed, as shown in Fig.~\ref{rlx}(a) and (b): 
(a) 9 Sm layers and 8 B$_6$ layers for the Sm-terminated slab,
(b) 9 Sm layers and 10 B$_6$ layers for the B$_6$-terminated slab.
In both cases, vacuum regions with 25 and 28 $\AA$ thickness, respectively, are 
considered in-between adjacent slabs.

For the DFT, we used both the full-potential linearized augmented-plane-wave (FLAPW) 
band method implemented in WIEN2k\cite{WIEN2k} and
the projector augmented wave (PAW) band method implemented in VASP.\cite{VASP}
In both cases, we have employed the generalized
gradient approximation (GGA) for the exchange-correlation potential.
Bond lengths and internal atomic positions were optimized using the conjugate-gradient algorithm
implemented in the VASP code
until the residual forces were less than 0.01 eV/$\AA$.
For the plane wave bases, energy cut-off of 300 eV was used.
{\bf
}
For the given relaxed structures obtained via VASP, 
we used both VASP and WIEN2k to cross-check DOSs and band structures.
In WIEN2k calculations, wave functions inside the muffin-tin (MT) spheres were expanded 
in the spherical harmonics up to $l_{max}$=10. 
The wave function in the interstitial region was expanded 
in the plane waves up to $K_{max} \times R_{MT}=7.0$ ($R_{MT}$: MT radius). 
$R_{MT}$'s of Sm and B are chosen as 2.5 a.u. and 1.5 a.u., respectively. 
$24\times 24\times 2$ {\bf k}-point mesh was used in the Brillouin zone. 
{\bf
}

The resulting relaxed structures are shown in Fig.~\ref{rlx}(a) and (b).
In both cases, the slabs are seen to be shrunk mainly within two topmost layers. 

\section{Influence of Sm $j=7/2$ bands on the states near E$_F$}

In our early investigation of bulk bands of SmB$_6$ using the DFT,\cite{CJ} we found that
Sm $j=7/2$ bands above E$_F$ are mixed somehow with the low-lying $f$-bands near E$_F$,
so as to produce the spurious charge densities.
Hence, to remove the influences of $j=7/2$ bands around E$_F$,
we performed the slab calculations with the 10 times stronger SOC,
as we did in the bulk case.
We have ascertained that the overall features of metallic surface states, such as 
number of FSs, relative FS sizes, and the hole or electron nature of each FS, 
are almost unchanged even with the stronger SOC for both terminations.
This result indicates that the effects of Sm $j=7/2$ bands on the metallic surface states 
 of SmB$_6$ near E$_F$ would be minor.

\section{Dirac state and Fermi surfaces centered at $\bar{M}$}

\begin{figure*}[b]
\begin{center}
\includegraphics[width=0.85\textwidth] {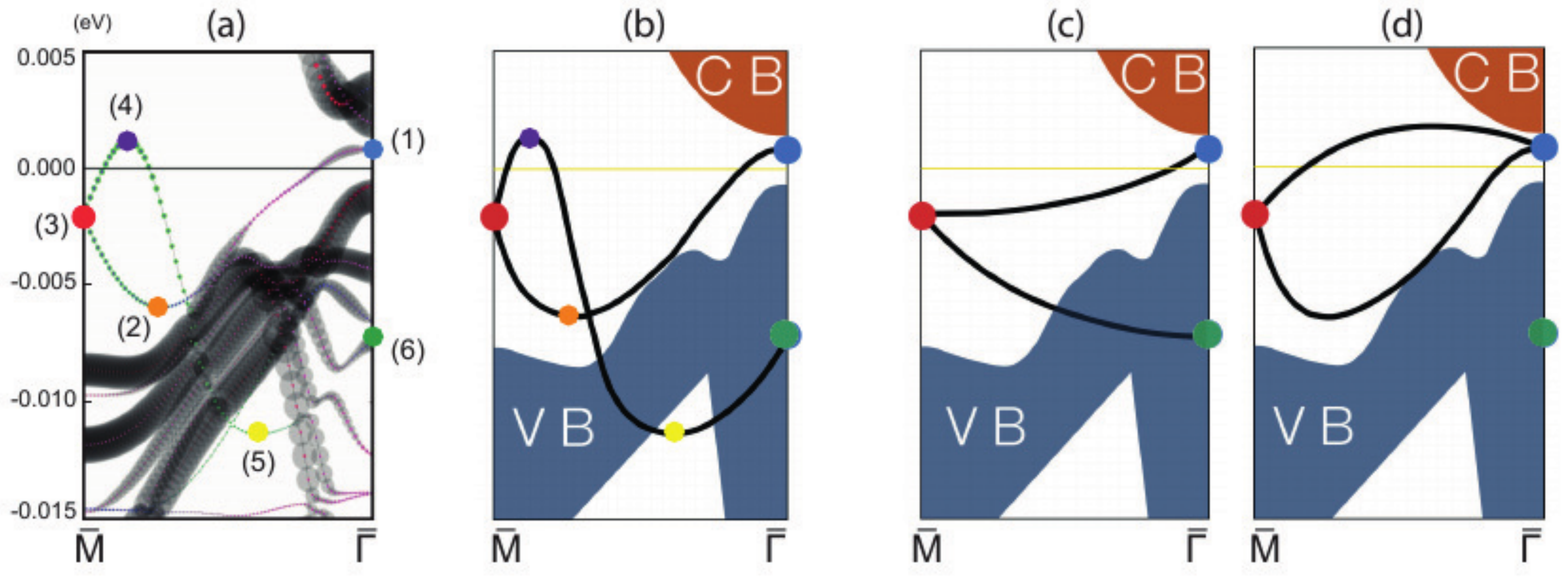}
\caption{
(Color online)
(a) The surface band structure along $\bar{M}$-$\bar{\Gamma}$ for the Sm-terminated SmB$_6$ slab.
Thick black circles represent the weight contributions from bulk bands.
To show the surface states distinctly, they are plotted with small colored circles that represent 
the weight contributions from surface B-2p states.
(b) Schematic band structure of (a), which shows the switching of two surface bands.
(c) Topologically nontrivial surface states in conventional topological insulators.
(d) Topologically trivial surface states in conventional band insulators.
CB and VB represent the projected bulk conduction and valence bands, respectively.
}
\label{ss_spectra}
\end{center}
\end{figure*}

Differently from $\bar{\Gamma}$ and $\bar{X}$-centered surface states in SmB$_6$,
$\bar{M}$-centered  surface states,
which emerge in the Sm-terminated SmB$_6$, have not been explored 
in the context of topological insulator.
In the topological insulators, Kramers degenerate pairs switch partners,
reflecting the change in the time reversal polarization.\cite{3D_TI}
Also, the corresponding surface states cross the Fermi level ($E_F$) an odd number of times 
between TRIM (time reversal invariant momentum) points having different 
time reversal polarizations, $-1$ or 1.
We checked the topological nature of the surface states at $\bar{M}$.
It is seen that the upper band at $\bar{M}$ first increases in energy above $E_F$ and then
decreases down so as to be connected to the band (6) at $\bar{\Gamma}$ below $E_F$,
while the lower band at $\bar{M}$ is connected to the band (1) at $\bar{\Gamma}$ above $E_F$,
as is schematically drawn in Fig.~\ref{ss_spectra}(b).
It shows typical features of topological insulators such as 
"switching partners" and "an odd number of E$_F$ crossing" [Fig.~\ref{ss_spectra}(c)],
which is clearly different from that of normal band insulators [Fig.~\ref{ss_spectra}(d)]. 
But the Dirac cone at $\bar{M}$ is seen to be reversed with respect to a normal Dirac cone 
in Fig.~\ref{ss_spectra}(c) that has monotonously decreasing or increasing dispersions.



\begin{figure*}[t]
\begin{center}
\includegraphics[width=0.7\textwidth]{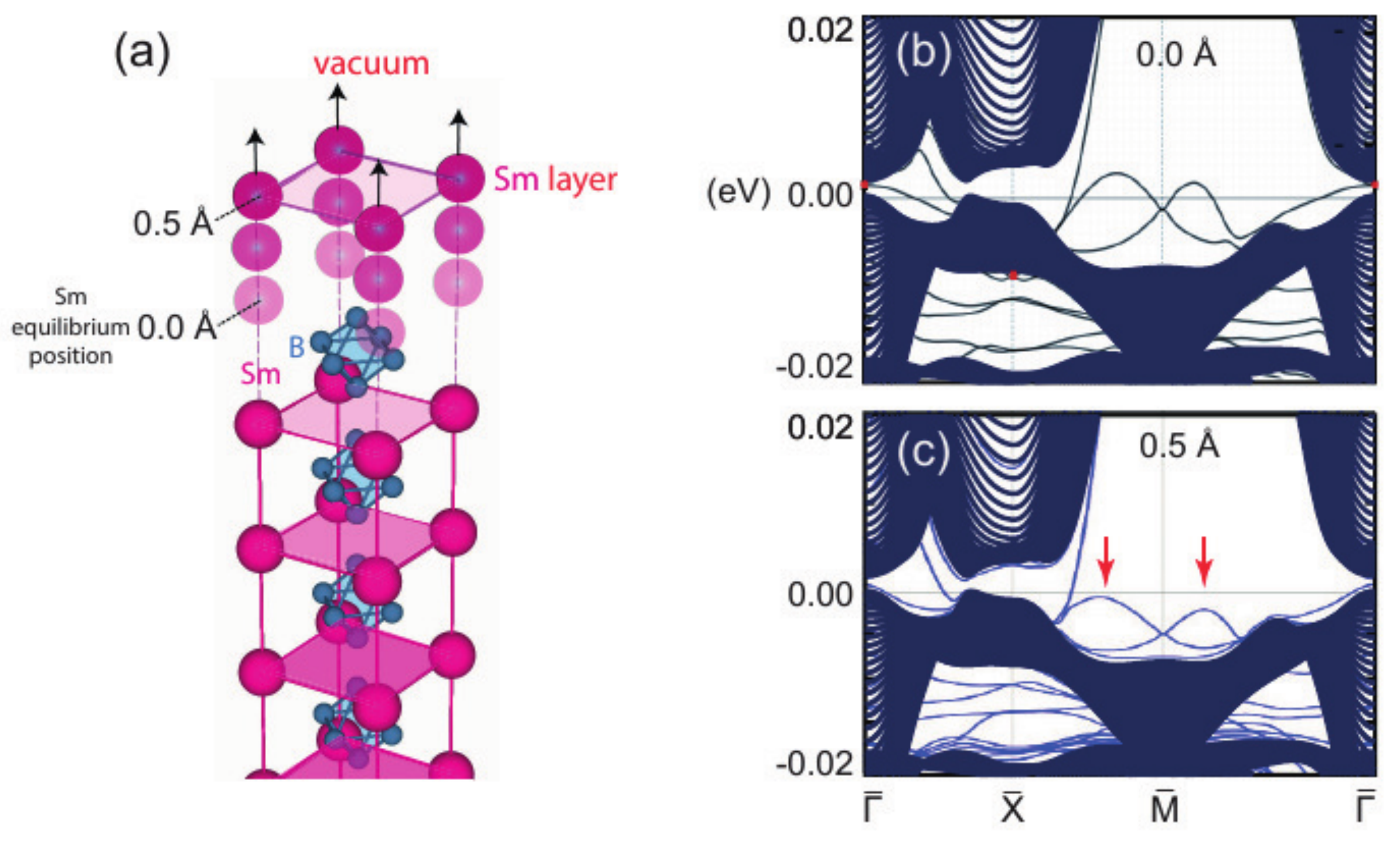}
\caption{
(Color online)
Variations of the topological surface states in the Sm-terminated SmB$_6$
with respect to a perturbation of lifting up the top Sm layer to vacuum side
from its equilibrium position by 0.5 $\AA$. 
A slab geometry of 11 Sm layers and 10 B$_6$ layers is used 
with 88 $\AA$ vacuum region in-between adjacent slabs.
(b),(c) Surface states before and after the perturbation. Fermi surfaces centered at $\bar{M}$
disappear due to the perturbation, while those centered at $\bar{\Gamma}$ and $\bar{X}$ are intact.
}
\label{perturbation}
\end{center}
\end{figure*}

To investigate the robustness of the Fermi surfaces, 
we examined the variation of the surface states with respect to the perturbation.
For that purpose, we considered the situation that the top Sm layer of the Sm-terminated slab 
was shifted up from its equilibrium position by 0.5 $\AA$ (see Fig.~\ref{perturbation}(a)).
As shown in Fig.~\ref{perturbation}(b) and (c),
due to this perturbation, two $\bar{M}$-centered Fermi surfaces disappear,
while $\bar{\Gamma}$ and $\bar{X}$-centered Fermi surfaces are intact. 
The relevant surface states at $\bar{M}$ become less dispersive and
the Dirac point becomes lower than before.
This feature suggests that $\bar{M}$-centered Fermi surfaces are likely to be more fragile 
than $\bar{\Gamma}$ and $\bar{X}$-centered Fermi surfaces.
Note that these two $\bar{M}$-centered Fermi surfaces are not compulsory for the topological 
insulating nature of SmB$_6$.

